\documentclass[twocolumn,aps,amsmath,amssymb,floatfix,superscriptaddress,prb,longbibliography]{revtex4-1}
\usepackage{graphicx}
\usepackage{dcolumn}
\usepackage{bm}
\usepackage[normalem]{ulem}
\usepackage{xspace}
\usepackage{hyperref}
\usepackage{cleveref}
\usepackage{soul}
\usepackage{color}
\usepackage[flushleft]{threeparttable}
\usepackage[utf8]{inputenc}
\usepackage{amsmath}

\begin{document}


\title{Resolving inconsistencies in Relativistic Thermodynamics with four-entropy and four-heat current}

\author{Biswajit Pandey}
\affiliation{Department of Physics, Visva-Bharati University, Santiniketan, Birbhum, 731235, India}
\email{biswap@visva-bharati.ac.in}

\begin{abstract}
  \noindent{\bf} We address long-standing inconsistencies in
  relativistic thermodynamics, particularly the ambiguities in
  temperature transformations and entropy evolution in moving or
  accelerating frames. Traditional approaches often lead to spurious
  entropy dissipation or generation in accelerating frames,
  contradicting the second law of thermodynamics. By introducing
  four-entropy and four-heat current, we resolve the problematic terms
  related to Lorentz factor dependence, eliminating unphysical entropy
  production. We also show that transformations like
  $T^{\prime}=\frac{T}{\gamma}$ and $T^{\prime}=T \gamma$ introduce an
  additional unphysical source of entropy dissipation or generation
  without real heat flow, driven by temperature gradients from
  relative motion. By adopting the covariant transformation
  $T^{\prime}=T$ and treating heat and entropy as four-vectors, we
  resolve these inconsistencies and offer a coherent framework for
  relativistic thermodynamics.
\end{abstract}
\maketitle

\noindent{\bf Introduction}\\
Classical thermodynamics was originally formulated in a
non-relativistic framework, where the motion of systems occurs at
velocities significantly lower than the speed of light, and time and
space are regarded as distinct and absolute. However, when systems
approach relativistic speeds or experience substantial acceleration,
the classical approach to thermodynamics encounters fundamental
conceptual difficulties.

One of the central challenges in relativistic thermodynamics is
understanding how physical quantities such as temperature, entropy,
and heat transform between inertial frames in relative motion. These
transformations must not only be consistent with the principles of
thermodynamics but also conform to special relativity, where space and
time are unified into spacetime, and the laws of physics remain
invariant across all inertial frames. The subject of relativistic
thermodynamics has been marked by controversy, largely due to
inconsistencies that emerge when classical thermodynamic concepts are
extended to systems traveling at relativistic speeds. A notable
example is the transformation of temperature, a topic that has fueled
considerable debate, with three primary viewpoints on the issue.

Einstein \cite{einstein07} and Planck \cite{planck08} proposed a
temperature transformation given by $T^{\prime}=\frac{T}{\gamma}$,
where $\gamma$ is the Lorentz factor, and $T$ and $T^{\prime}$
represent the temperatures in the rest frame and the moving frame,
respectively. This formulation suggests that objects in motion should
appear colder. In contrast, Ott \cite{ott63} and Arzelies
\cite{arzelies65} arrived at a different result, $T^{\prime}=T
\gamma$, which implies that moving bodies should appear hotter. A
third viewpoint, introduced by Landsberg \cite{landsberg67,
  landsbergJohns67}, asserts that temperature is a Lorentz-invariant
quantity, with the relation $T^{\prime}=T$. While each of these
perspectives has garnered support through various studies, no
definitive consensus has been reached. These inconsistencies pose
challenges not only to the principles of relativity but also to the
second law of thermodynamics.

Several previous studies have introduced a covariant formulation of
thermodynamics by employing the concept of four-temperature
\cite{dantzig39, vankampen68, israel81, nakamura06,
  wu09}. Nevertheless, the debate concerning the relativistic
transformation of temperature remains unresolved. In this work, we
investigate how the concept of four-entropy, in conjunction with the
four-vector heat current, can provide a unified and consistent
approach to relativistic thermodynamics. By addressing ambiguities in
the evolution of entropy and the transformation of temperature, we
align relativistic effects with the laws of thermodynamics, offering a
robust framework for analyzing thermodynamic systems in relativistic
regimes.\\

\noindent{\bf Inconsistencies in the classical entropy evolution
  equation}\\ 
In classical thermodynamics, entropy $S$ represents the degree of
disorder within a system. For a closed system, changes in entropy $dS$
are governed by the second law of thermodynamics, which states that
$dS=\frac{dQ}{T}$, where $dQ$ is the infinitesimal amount of heat
added to the system and $T$ is its temperature. Entropy always
increases in irreversible processes while it remains constant in
reversible processes.

The second law of thermodynamics guarantees that in any process, the
entropy of an isolated system cannot spontaneously decrease. This
principle functions effectively in non-relativistic scenarios, where
time is absolute and reference frames do not influence the measurement
of thermodynamic quantities. However, complications emerge when
extending this concept to relativistic contexts, especially when a
system is observed from a moving or accelerating frame.

Consider a box containing a monatomic gas in an inertial frame at
rest. The gas occupies a volume $V$ and consists of $N$ atoms at
temperature $T$ in this rest frame. The entropy $S$ of the system in
the rest frame can be determined using the Sackur-Tetrode equation
\cite{sackur12, tetrode12}, which provides an expression for the
entropy of an ideal monatomic gas based on its temperature, volume,
and the number of particles. According to this equation the entropy
$S$ in the rest frame is given by,
\begin{equation}
\begin{split}
S=N k_B\left[\ln\left\{\left(\frac{V}{N}\right)\left(\frac{2 \pi m k_B
      T}{h^2}\right)^\frac{3}{2}\right\}+\frac{5}{2}\right].
\label{eq:ent}
\end{split}
\end{equation}
Here, $m$ represents the mass of a single atom, and $h$ denotes
Planck's constant. Now, consider a second inertial frame moving with a
relativistic velocity $v$ along the positive x-axis with respect to
the rest frame. We would like to determine how the entropy of the gas
transforms between these two frames, taking into account the
relativistic effects associated with high velocities. Understanding
this transformation is key to reconciling thermodynamic principles
with the framework of special relativity. Using
$T^{\prime}=\frac{T}{\gamma}$ and $V^{\prime}=\frac{V}{\gamma}$, the
entropy $S^{\prime}$ in the moving frame is related to the entropy $S$
in the rest frame as follows,
\begin{equation}
S^{\prime}=S-\frac{5}{2} N k_{B} \ln \gamma,
\label{eq:ent_trans}
\end{equation}
where $k_B$ is the Boltzmann constant. The $\ln \gamma$ term persists
regardless of which temperature transformation is employed. Notably,
the $\gamma$ factor in the Sackur-Tetrode equation in the moving frame
does not cancel out for any temperature transformation. The sign
associated with the $\ln \gamma$ term will be positive for the
transformation $T^{\prime}=T \gamma$ and negative for the
transformation $T^{\prime}=T$.

Some may propose the use of a relativistic version of the
Sackur-Tetrode equation to express the entropy in the moving
frame. Intriguingly, the entropy $S^{\prime}$ in the moving frame
still remains related to the rest frame entropy $S$ through a
logarithmic factor involving the Lorentz factor $\gamma$
\citep{taff68}. The presence of the $\ln \gamma$ term in the entropy
transformation underscores a fundamental connection between the
principles of relativity and thermodynamics.

We now investigate how this term influences the validity of the second
law of thermodynamics. For simplicity, let us express the entropy
transformation in \autoref{eq:ent_trans} as $S^{\prime}=S-\ln \gamma$,
neglecting the coefficient of $\ln \gamma$ since it varies depending
on the type of gas or system under consideration. Taking the
derivative with respect to the proper time $\tau$ in the moving frame, we
obtain,
\begin{equation}
\frac{dS^{\prime}}{d\tau}=\gamma \frac{dS}{dt}-\gamma^3
\frac{\vec{v}\cdot\vec{a}}{c^2},
\label{eq:ent_evol}
\end{equation}
where $a$ is the acceleration of the moving frame and $c$ is the speed
of light in vacuum. If the acceleration $a=0$ or
$\vec{v}\perp\vec{a}$,, only the first term in \autoref{eq:ent_evol}
remains. The first term guarantees that $\frac{dS^{\prime}}{d\tau}
\geq 0$, given that $\frac{dS}{dt}\geq 0$ in the rest
frame. Consequently, the second law of thermodynamics is upheld in
both frames, provided the moving frame is either not accelerating or
its acceleration is perpendicular to its velocity. It is worth noting
that while special relativity primarily deals with inertial frames, it
can still handle accelerating frames by locally approximating them as
a series of infinitesimal inertial frames.

We encounter a conceptual difficulty with the accelerating frames in
the present context. The second term in this equation implies that
entropy is either generated or dissipated in the moving frame purely
due to acceleration, even in the absence of physical heat transfer or
irreversible processes. This indicates that in an accelerating frame,
entropy changes as a result of the relative motion of the observer,
even if the system remains in equilibrium in its rest frame.

This situation presents a significant inconsistency. According to the
second law of thermodynamics, entropy should only increase as a result
of irreversible processes such as heat transfer or friction. However,
the dissipation term in \autoref{eq:ent_evol} suggests that entropy
can decrease or increase due to the acceleration of the observer,
which is unphysical. This generation of entropy does not correspond to
any actual heat flow or dissipation mechanism within the system. It
arises purely from the kinematic effects of acceleration, independent
of any real thermodynamic process.

Moreover, even at small accelerations, the second term in
\autoref{eq:ent_evol} can dominate the first at ultra-relativistic
velocities, potentially leading to a net loss of entropy in the moving
frame. Such a scenario could result in
$\frac{dS^{\prime}}{d\tau} < 0$, signaling a breakdown of the
second law of thermodynamics. However, this apparent contradiction
arises from treating entropy as a scalar quantity. The conventional
understanding of entropy requires modification to incorporate the
relative motion between the system and the observer. In relativistic
contexts, entropy becomes frame-dependent, with its rate of change
influenced by both the velocity and acceleration of the observer. This
underscores the necessity for a relativistic extension of classical
thermodynamics, in which entropy is no longer a mere scalar but is
affected by the relative motion of the observer.

It may be tempting to connect the entropy dissipation observed in the
accelerating frame with Unruh radiation \citep{davies75, unruh76}
detected by an accelerating observer. Both effects are indeed linked
to acceleration. However, this connection is not tenable because the
entropy dissipation in the classical thermodynamic context stems from
a dependence on the Lorentz factor, resulting in an unphysical
decrease in entropy without any real heat flow. In contrast, Unruh
radiation represents a genuine quantum phenomenon, where an
accelerating observer perceives a thermal bath of particles, leading
to actual entropy production through energy exchange with this
radiation. The classical entropy dissipation is problematic because it
does not involve any real thermodynamic process, whereas Unruh
radiation reflects physical interactions rooted in quantum field
theory. This distinction underscores that the two phenomena are
fundamentally different.\\

\noindent{\bf The four-entropy in relativistic thermodynamics}\\
In this work, we treat entropy as a four-vector rather than as a
scalar. We define the four-entropy as $S^{\mu}=s U^{\mu}$, where $s$
is the entropy density (a scalar quantity) in the rest frame of the
system, and $U^{\mu}=(\gamma c, \gamma \vec{v})$ is the
four-velocity. In this framework, entropy is viewed as part of a flow
through space and time, with the four-velocity $U^{\mu}$ determining
how entropy behaves across different frames.

The four-entropy transforms covariantly under Lorentz transformations,
ensuring consistency across all reference frames, including
accelerating ones. When performing a Lorentz transformation to a
reference frame moving with velocity $v$ along the positive x-axis
relative to the rest frame, the four-entropy components transform as
$S^{\prime \mu}=\Lambda_{\nu}^{\mu} S^{\nu}$, where
$\Lambda_{\nu}^{\mu}$ is the Lorentz transformation matrix. This
formulation guarantees that the second law of thermodynamics, which
states that the total entropy of an isolated system must either
increase or remain constant, holds in any reference frame. Crucially,
the transformation of four-entropy is consistent with both
relativistic kinematics and thermodynamics, providing a coherent way
to reconcile the two in relativistic contexts.

In relativistic thermodynamics, the second law takes the generalized form
\begin{equation}
\partial_{\mu} S^{\mu} \geq 0,
\label{eq:entre}
\end{equation}
where $\partial_{\mu}$ is the four-divergence operator. This
formulation of the second law of thermodynamics is Lorentz-invariant,
meaning it holds in all inertial reference frames and avoids the
contradictions that arise when entropy is treated as a scalar in a
relativistic context. The non-negative four-divergence of the
four-entropy ensures that the rate of entropy production remains
consistent across all frames, including accelerating ones. While we
introduced the concept of four-entropy within the framework of
inertial frames, it can also be extended to non-inertial frames. We
provide a detailed discussion on the generalized form of the second
law of thermodynamics in the proposed four-vector framework and
discuss its implications in the Appendix (\autoref{sec:appen}).\\

\noindent{\bf Resolution of classical entropy evolution with
  four-entropy}\\ Let us resolve the contradiction in the classical
entropy evolution equation (\autoref{eq:ent_evol}) using the concept
of four-entropy.

In our example, the primed frame moves along the positive
x-direction. Therefore, the Lorentz transformation matrix is given by
\begin{equation}
\Lambda_{\nu}^{\mu} = \begin{bmatrix} 
	\gamma & -\gamma \frac{v}{c} & 0 & 0 \\
	-\gamma \frac{v}{c} & \gamma & 0 & 0\\
        0 & 0 & 1 & 0\\
        0 & 0 & 0 & 1\\
	\end{bmatrix}.
\label{eq:lorentz}
\end{equation}
The four-entropy of the system in the rest frame is
$S^{\mu}=(sc,0,0,0)$. Applying the Lorentz transformation, the
four-entropy in the primed frame becomes $S^{\prime \mu}=(\gamma s c,
-\gamma s v,0,0)$.

We defined four-entropy as $S^{\mu}=s U^{\mu}= s (\gamma c, \gamma
\vec{v})$. Assuming the entropy density $s$ is uniform in space and
time, we now calculate the four-divergence of the four-entropy, which
is given by,
\begin{equation}
\partial_{\mu} S^{\mu}=s \gamma^3 \frac{\vec{v}\cdot\vec{a}}{c^2}+ \nabla \cdot ({ s \gamma \vec{v}}),
\label{eq:generalized}
\end{equation}

In the rest frame of the system, we have $\gamma=1$, $\nabla \cdot
\vec{v}=0$, and $\vec{a}=0$. We assume $s=$ constant in the rest frame
of the system. Therefore, $\partial_{\mu} S^{\mu}=0$ in the rest frame
of the system. The four-divergence of four-entropy in the primed frame
must also equal zero. Thus, from \autoref{eq:generalized}, we have
$\nabla \cdot (\gamma \vec{v})=-\gamma^3
\frac{\vec{v}\cdot\vec{a}}{c^2}$. This shows that the problematic
dissipation term in the classical entropy evolution equation
originates from the non-zero entropy flux $\nabla \cdot (\gamma s
\vec{v})$ in the primed frame. Here, the entropy current density $s
\vec{v}$ is scaled by a factor $\gamma$ in the moving frame, and the
resulting entropy flux comes from the spatial derivative of the
entropy current density. If the primed frame moves with an uniform
velocity, then $\nabla \cdot (\gamma \vec{v})=0$, implying that there
would be no entropy dissipation or generation in the primed frame due
to the relative motion between the system and the observer. This
conclusion also holds when $\vec{v}\cdot\vec{a}=0$, implying that
there is no divergence of the entropy current density when the
velocity and acceleration are orthogonal (as in circular motion).

In classical thermodynamics, entropy flux is typically associated with
processes such as heat conduction or particle diffusion. However, in a
relativistic context, entropy flux includes an additional contribution
due to the relative motion between the system and the observer. The
dissipation term $\gamma^3\frac{\vec{v}\cdot\vec{a}}{c^2}$, which
arises in the relativistic regime, stems from the time component of
the four-divergence and is tied to the non-zero flux of entropy
current density in the accelerating frames. The four-entropy formalism
thus provides a consistent, Lorentz-invariant description of entropy
flow, thereby avoiding unphysical outcomes like entropy dissipation or
generation purely due to acceleration, while consistently upholding
the second law.\\

\noindent{\bf Four-vector heat current and its relation to
  four-entropy}\\In non-relativistic thermodynamics, heat flow is
described as the transfer of thermal energy within a system. It is
traditionally treated as a scalar quantity, representing the amount of
heat transferred per unit area per unit time. In classical systems,
heat flow is driven by temperature gradients and is governed by
Fourier’s law of heat conduction, which states that the heat transfer
rate is proportional to the negative of the temperature gradient. It
is expressed as,
\begin{equation}
q=-k \nabla T,
\label{eq:fourier}
\end{equation}
where $q$ is the heat flux density, $k$ is the conductivity and
$\nabla T$ represents the temperature gradient.

In the present framework, heat is no longer treated as a scalar but as
part of the energy-momentum tensor $T^{\mu\nu}$, which encodes the
energy density, momentum density, and stress of the system.  We define
heat current $Q^{\mu}$ as a four-vector as,
\begin{equation}
Q^{\mu}=\gamma \left(c q_{0}, \vec{q}\right),
\label{eq:heatcurrent}
\end{equation}
where $q_{0}$ represents the rest frame heat energy density. The time
component of $Q^{\mu}$ describes the relativistically boosted total
energy flux perceived by a moving observer. The space components
represent spatial heat flux, which describes the directional flow of
heat per unit area, per unit time. The time component can also be
viewed as the $T^{00}$ component in the energy momentum tensor. The
spatial components of the heat current $Q^{\mu}$ correspond to the
energy flux ($T^{0\nu}$ components where $\nu=1,2,3$) in the
energy-momentum tensor, describing how thermal energy flows through
spacetime. This ensures that both the magnitude and direction of heat
transfer adjust appropriately when observed from a moving frame,
taking into account effects like time dilation and Lorentz
contraction. In a moving frame, the heat current $Q^{\mu}$ evolves due
to acceleration, leading to entropy dissipation or generation. We will
show that this explains the appearance of the dissipation term in the
entropy evolution equation (\autoref{eq:ent_evol}).

The change in four-entropy of the system can be expressed in terms of
the heat transfer as,
\begin{equation}
dS^{\mu}=\frac{dQ^{\mu}}{T},
\label{eq:entheat}
\end{equation}
where $dS^{\mu}$ represents the infinitesimal change in the
four-entropy, $dQ^{\mu}$ is the infinitesimal heat transfer as a
four-vector, and $T$ is the temperature of the system. This equation
provides the relativistic version of the heat-entropy relationship,
where both heat and entropy are treated as four-vectors.

In the relativistic framework, we can write
$S^{\mu}=\frac{Q^{\mu}}{T}$, which holds for reversible processes
where heat and entropy flow are directly linked. This assumes the
system is in local thermal equilibrium. We use this relation as a
natural extension of the classical thermodynamic equation into
four-dimensional spacetime, with both entropy and heat described as
four-vectors. The four-divergence of the entropy current is
\begin{equation}
  \partial_{\mu}
  S^{\mu}=\partial_{\mu}\left(\frac{Q^{\mu}}{T}\right)\\ =\frac{1}{T}
  \partial_{\mu} Q^{\mu}+ Q^{\mu} \partial_{\mu}\left(\frac{1}{T}\right)
\label{eq:4div_ent1}
\end{equation}
The first term represents the rate of heat flow per unit temperature,
while the second term reflects the impact of temperature gradients on
entropy production. This second term is associated with irreversible
processes like heat conduction or diffusion. Different relativistic
temperature transformations will affect this term in distinct ways.

The four-divergence of the entropy current, $\partial_{\mu}S^{\mu}$,
represents the rate of entropy production. If we assume the heat
current $Q^{\mu}$ satisfies the conservation equation $\partial_{\mu}
Q^{\mu} = 0$, then entropy production arises solely due to temperature
gradients.
\begin{equation}
  \partial_{\mu} S^{\mu}= Q^{\mu} \partial_{\mu}\left(\frac{1}{T}\right)
\label{eq:4div_ent2}
\end{equation}
This indicates that even when heat is conserved, entropy can still be
generated in the presence of temperature gradients. On the other hand,
if $\partial_{\mu} Q^{\mu} \neq 0$ (i.e., heat flow is not conserved
due to dissipation), the first term $\frac{1}{T} \partial_{\mu}
Q^{\mu}$ will also contribute to entropy production. Thus, the
four-divergence of the entropy current accounts for both heat flow and
temperature gradients. When acceleration causes a non-zero divergence
of the heat current, it introduces non-equilibrium processes that
generate entropy. The dissipation term $\gamma^3
\frac{\vec{v}\cdot\vec{a}}{c^2}$ in the classical entropy evolution
equation can be interpreted as arising from this non-zero divergence
in the presence of acceleration, demonstrating the consistency between
relativistic thermodynamics and classical entropy evolution.

The four-divergence of the heat current and the entropy current become
equal when the second term, $Q^{\mu} \partial_{\mu}(\frac{1}{T})$,
vanishes under specific conditions. If the temperature $T$ is uniform
across spacetime, with no spatial or temporal gradients, then
$\partial_{\mu}(\frac{1}{T})=0$. However, if $T$ varies with space and
time, this term introduces additional entropy production or
dissipation due to the temperature gradient. When the heat current is
conserved (i.e., there is no net flow of heat energy), the
four-divergence of the entropy current is also zero.

In our case, the first term in \autoref{eq:generalized}, which
represents dissipation in the accelerating frame, must be balanced by
the second term, involving the divergence of the bulk velocity of the
system and the gradient of the Lorentz factor. To explore this with
four-heat current, we first calculate the four-divergence of
$Q^{\mu}$, assuming the temperature is constant in spacetime. We will
later consider the effects of a temperature that varies in space and
time.

The four-divergence of the heat current $Q^{\mu}$ is given by,
\begin{equation}
  \partial_{\mu} Q^{\mu}= \gamma^3
  \frac{\vec{v}\cdot\vec{a}}{c^2} q_{0}+\gamma
  \frac{\partial q_{0}}{\partial t}+  \nabla \cdot (\gamma \vec{q})
\label{eq:4div_heat}
\end{equation}
The first term signifies how the energy density felt by the observer
changes as they accelerate. It would be zero if there is no
acceleration. $\frac{\partial q_{0}}{\partial t}$ in the second term
represents time rate of change of the energy density in the rest frame
of the system. This would be zero in our case as the energy density
$q_0$ remains constant under equilibrium condition. Therefore, the
time derivative of the time component of heat current is entirely
described by the first term. The third term, $ \nabla \cdot (\gamma
\vec{q})$, represents the divergence of the heat current density,
which measures the net heat flow out of a given volume. This term also
involves the spatial variation of the Lorentz factor. 

We have $\partial_{\mu} Q^{\mu} = 0$ in the rest frame of the
system. In the accelerating frame, the gradient of $\gamma$ would
affect how an observer in an accelerating frame measures the heat
energy density or heat flux, but it does not imply that heat is being
produced out of nothing or that entropy is being generated in
violation of the second law of thermodynamics. The divergence of the
heat current density will be balanced by the combined effects of real
heat transfer, spatial heat flow, the time dependence of Lorentz
factor and the gradient of $\gamma$.\\

\noindent{\bf Ambiguities in temperature transformations}\\ The heat
current $Q^{\mu}$, which describes the flow of thermal energy, evolves
in spacetime, leading to entropy production in the accelerating frame.
We interpret the problematic dissipation or production term, $\gamma^3
\frac{\vec{v}\cdot\vec{a}}{c^2}$, in the classical entropy evolution
equation as a result of the non-zero divergence of the four-entropy
current or four-heat current in a relativistic, accelerating
system. This term represents the additional entropy dissipation or
generation caused by heat flow due to acceleration. It arises from the
non-equilibrium conditions introduced by the accelerating frame.

We now turn our attention to the ambiguities that may arise in this
framework due to different temperature transformations. Entropy
production is closely linked to irreversible processes, such as heat
flow across temperature gradients. When there are temperature
gradients, heat flows naturally from hotter to colder regions,
generating entropy as required by the second law of thermodynamics. If
we accept the temperature transformation
$T^{\prime}=\frac{T}{\gamma}$, an observer moving relative to the
system would observe a decrease in temperature. This
velocity-dependent change can introduce spatial or temporal
temperature gradients. Since $\gamma$ depends on time due to
acceleration, the observer perceives a time-dependent temperature
gradient that drives heat flow. These gradients can result in entropy
dissipation, as the system may appear to be out of thermal equilibrium
in the accelerating frame due to the varying temperature across space
and time. 

Conversely, if we consider the temperature transformation
$T^{\prime}=T \gamma$, the increase in temperature observed in the
moving frame can also create a temperature gradient. This gradient
would drive heat flow, leading to entropy generation.

In both cases, entropy dissipation or generation occurs because the
transformations create temperature gradients, which result in heat
flow and irreversible processes that produce entropy. An accelerating
observer perceives these gradients, even if the system remains in
equilibrium in the rest frame. Thus, for these temperature
transformations, both terms in \autoref{eq:4div_ent1} contribute to
entropy dissipation or generation. The second term indicates that
entropy is produced or dissipated due to variations in the Lorentz
factor across space and time.  However, its physical origin is
distinctly different from the first term. 

The time-dependent Lorentz factor, $\gamma$, indeed contributes to
entropy generation or dissipation, as shown in the case of a non-zero
divergence of the four-entropy or four-heat current. However, the main
problem with the transformations $T^{\prime}=\frac{T}{\gamma}$ and
$T^{\prime}=T \gamma$ is that they create temperature variations
across spacetime. As velocity changes, different temperatures are
observed in different regions or at different times, leading to
temperature gradients. These gradients directly cause entropy
dissipation or generation without actual heat flow in the rest frame,
which violates the second law of thermodynamics. These transformations
suggest temperature shifts solely due to relative velocity, lacking
proper thermodynamic justification. From a thermodynamic
standpoint, temperature should not change purely due to the motion of
the observer unless there is a genuine process, such as heat transfer
or work being done on the system. Thus, even though heat and entropy
flux can be described using four-vectors in an accelerating frame, the
issue with the transformations $T^{\prime}=\frac{T}{\gamma}$ and
$T^{\prime}=T \gamma$ remains problematic.

The entropy transformation in \autoref{eq:ent_trans} introduces a term
of $\frac{5}{2}N k_{B} \ln \gamma$ in the moving frame. Part of this
term, $N k_{B} \ln \gamma$, arises from transformation of volume,
while the remaining $\frac{3}{2} N k_{B} \ln \gamma$ results from the
velocity-dependent temperature transformation
$T^{\prime}=\frac{T}{\gamma}$. This additional term, generated by the
velocity-dependent temperature transformation, causes unphysical
entropy production in the accelerating frame, thereby violating the
second law of thermodynamics. It is important to note that this extra
contribution does not appear under the invariant temperature
transformation $T^{\prime}=T$, which prevents any contradiction with
the second law of thermodynamics.

The temperature of the cosmic microwave background (CMB) radiation
serves as an interesting comparison to temperature transformations in
relativistic thermodynamics. The CMB temperature is frame-dependent,
and the dipole anisotropy observed in the CMB is a result of our local
motion relative to the radiation field, producing a Doppler shift that
alters the observed temperature in different directions. This Doppler
shift in photon frequencies is tied to real, measurable physical
processes, and it reflects a well-understood and consistent
thermodynamic behavior. On the other hand, temperature transformations
in relativistic thermodynamics, such as $T^{\prime}=\frac{T}{\gamma}$
and $T^{\prime}=T \gamma$, introduce conceptual difficulties. These
transformations suggest a frame-dependent temperature that could imply
entropy dissipations or generation due to changes in the Lorentz
factor $\gamma$, even when no physical heat transfer occur, posing a
challenge to the second law of thermodynamics in accelerating
frames.\\

\noindent{\bf Why $T^{\prime}=T$ is preferred?}\\ The temperature of a
system, as a fundamental thermodynamic quantity, should not be
affected by purely kinematic effects. The transformations
$T^{\prime}=\frac{T}{\gamma}$ and $T^{\prime}=T \gamma$, which imply
that temperature varies with the relative velocity of the observer,
introduce entropy production without any corresponding heat flow, thus
violating the second law of thermodynamics. Ideally, relativistic
temperature should remain invariant unless a physical process
justifies its change. This perspective rests on the idea that
temperature, as a measure of internal thermal energy, should remain
unaffected by a Lorentz boost, as it depends solely on the random
microscopic motion of particles relative to each other and not on the
bulk motion of the system relative to an observer. The increase in
total energy observed in a moving frame arises from the bulk kinetic
energy associated with the velocity of the system relative to the
observer, not from internal thermal energy. Thus, while the total
energy increases, temperature, reflecting the internal energy, does
not need to change, making $T^{\prime}=T$ a consistent transformation
for thermodynamic systems in equilibrium.

The transformation $T^{\prime}=T$, where temperature remains invariant
under Lorentz transformations, preserves the correct scalar nature of
temperature. This approach ensures that entropy generation remains
linked to real thermodynamic processes, such as heat flow, as
described by the four-vector heat current $Q^{\mu}$. By treating
temperature as a Lorentz-invariant scalar, the transformation
$T^{\prime}=T$ avoids the inconsistencies seen with other
transformations, ensuring that no artificial temperature gradients or
unphysical entropy changes occur in the moving frame. This approach
aligns with the second law of thermodynamics, preventing entropy from
being generated or dissipated due to mere relative motion, and
ensuring that temperature remains consistent across different frames
without violating fundamental thermodynamic principles.\\

\noindent{\bf Conclusions}\\
Entropy is traditionally treated as a scalar quantity. For the
relationship $dS=\frac{dQ}{T}$ to maintain invariance, both heat $dQ$
and temperature $T$ must transform consistently. However, in this
work, we approach heat and entropy differently by defining them as
four-vectors that adhere to the principles of special relativity.

In an accelerating frame, the observed heat flux arises not from
direct heat transfer in the rest frame, but rather from the
transformation of the thermodynamic state of the system. Likewise, the
entropy flux illustrates how entropy changes as it transforms between
different reference frames. This treatment ensures that the fluxes
remain consistent with the four-divergence of both the heat current
and the entropy current in the accelerating frame, honouring the
second law of thermodynamics as well as the principles of relativity.

The classical equation for entropy evolution led to spurious entropy
dissipation or generation in the accelerating frame due to the
time-dependent Lorentz factor. In the four-vector formalism, spurious
entropy production arises from the time derivative and the gradient of
the Lorentz factor $\gamma$. These terms are counterbalanced by the
flux of four-entropy and four-heat current density in the accelerating
frame. Transformations such as $T^{\prime}=\frac{T}{\gamma}$ and
$T^{\prime}=T \gamma$ introduce an unphysical source of entropy
dissipation and generation in the accelerating frame, resulting from
the emergence of temperature gradients in both space and time.
Although this entropy production is also related to the time
derivative and gradient of the Lorentz factor $\gamma$, the heat flow
and entropy production resulting from temperature gradients in the
accelerating frame render it unphysical and inconsistent with the
second law of thermodynamics. We can eliminate this unphysical entropy
production by applying the covariant transformation $T^{\prime}=T$.
By adopting this transformation and treating heat and entropy as
four-vectors, we effectively resolve the inconsistencies present in
relativistic thermodynamics.

\bigskip
\noindent {\bf Data availability}\\
\noindent No datasets were generated or analyzed during the current study.

\bigskip
\noindent {\bf Acknowledgements}\\
\noindent The author thanks IUCAA, Pune, India for providing support through associateship programme.



\section{\bf Appendix}
\label{sec:appen}

\noindent{\bf Causality and Entropy: The relativistic roots of the
  second law of thermodynamics}\\
The second law of thermodynamics is one of the most fundamental
principles governing physical systems. It states that the total
entropy of an isolated system never decreases over time, defining a
natural direction to the flow of time, often referred to as the
``arrow of time''. This law governs a vast range of processes,
starting from molecular diffusion and star cooling to black hole
formation and the expansion of the universe. However, despite its
foundational importance, the origin of this law is typically treated
as axiomatic rather than derived from more basic physical principles.

In non-relativistic thermodynamics, the second law is often stated as
$\frac{dS}{dt} \geq 0$, indicating that entropy can only remain
constant or increase over time. However, in the relativistic context,
both time and space are observer-dependent, raising questions about
how entropy transforms across frames in relative motion. A key
challenge in relativistic thermodynamics is to determine how
quantities like temperature, entropy, and heat transform between
inertial frames moving with respect to each other. These
transformations must preserve thermodynamic principles while aligning
with special relativity. Extending classical thermodynamics to
relativistic systems has sparked considerable debate due to apparent
inconsistencies \citep{einstein07, planck08, ott63, arzelies65,
  landsberg67, landsbergJohns67, taff68, dantzig39, vankampen68,
  israel81, nakamura06, wu09}. This work, shows that these
inconsistencies can be resolved by utilizing four-entropy and
four-heat current and by adopting an invariant temperature
transformation, $T^{\prime}=T$. The assumed form of the second law of
thermodynamics $\partial_{\mu} S^{\mu}\geq 0$ in the proposed
four-vector framework justifies some explanation.

In this section, we would like to understand the second law of
thermodynamics within the framework of special relativity. Can we
justify this law through the principles of special relativity?
Entropy, often described as a measure of disorder, tends toward states
of maximum unpredictability or randomness. Yet, why must entropy
always increase? Special relativity, with its focus on causality and
invariant light cones, provides valuable insight into this
question. We demonstrate that the causal boundaries defined by
spacetime structure fundamentally shape the behaviour of entropy,
making the second law a natural and inevitable outcome. Using the
concept of four-entropy and its behaviour under Lorentz
transformation, we can describe the spatial flow and time evolution of
entropy in a self-consistent relativistic framework. In this
framework, the second law arises naturally from the structure of
spacetime, with entropy irreversibility enforced by causal
constraints.\\

\noindent{\bf Four-entropy and the generalized second law of thermodynamics}\\
In classical thermodynamics, entropy $S$ is an extensive scalar
quantity, representing the level of microscopic disorder within a
system.  However, in special relativity, time and space are treated
equivalently, implying that scalar entropy $S$ can not depend solely
on time. We must generalize the concept of entropy where it becomes a
component of a four-vector, the four-entropy $S^{\mu}$, which
incorporates both temporal and spatial components aligned along the
axes of spacetime. We define the four-entropy vector as $S^{\mu}= (s
\gamma c, s \gamma \vec{v})$, where, $s$ denotes the local entropy
density in the rest frame, $\vec{v}$ is the velocity vector of an
element of the system or the system itself, $c$ is the speed of light
in vacuum, and $\gamma$ is the Lorentz factor, accounting for time
dilation effects when the system or the subsystems move with respect
to an observer.  If $s_i$ and $\vec{v_{i}}$ are respectively the
entropy density and velocity of the $i^{th}$ subsystem within a system
then the four entropy of the system becomes $S^{\mu}= (\sum_{i} s_i
\gamma_i c, \sum_{i} s_i \gamma_i \vec{v_i})$. In any other inertial
frame moving at velocity $v$ relative to the system, the components of
the four-entropy $S^\mu$ transform as $S^{\prime
  \mu}=\Lambda_{\nu}^{\mu} S^{\nu}$, where $\Lambda_{\nu}^{\mu}$ is
the Lorentz transformation matrix. The spatial entropy flux in any
moving frame accounts both for the internal flows within the system
and for the relative motion of the observer.

The significance of describing entropy in this form lies in the fact
that the four-entropy encompasses both entropy density and entropy
flux. The entropy flux is a measure of the directional transport of
entropy within the system or with respect to a moving observer. By
studying the four-divergence of $S^{\mu}$, we can analyze how entropy
evolves over time and space in a way that aligns with relativistic
principles.

The generalized second law in this context is given by $\partial_{\mu}
S^{\mu} \geq 0$, where $\partial_{\mu} =\frac{\partial}{\partial
  X^{\mu}}=(\frac{1}{c}\frac{\partial}{\partial t}, \nabla)$ is the
derivative with respect to the contravariant components of spacetime
four-vector ($X^{\mu}$) and $\partial_{\mu} S^{\mu}$ is the
four-divergence of the four-entropy.

In relativity, $\partial_{\mu} S^{\mu}$ is Lorentz invariant, ensuring
that the generalized second law holds across all frames. This
invariance allows us to confirm that the second law of thermodynamics
applies consistently across all inertial and non-inertial frames.\\

\noindent{\bf Four-divergence of four-entropy in different
  frames}\\
In relativistic thermodynamics, the four-divergence of the
four-entropy current encompasses various sources of entropy production
and dissipation in the system.

Let us consider a system where the entropy density $s$ is potentially
non-uniform in space and time. We write the general expression for the
four-divergence of the four-entropy of the system as,

\begin{equation}
 \begin{split}
  \partial_{\mu} S^{\mu}  &=\frac{\partial (s \gamma)}{\partial t} +
  \nabla \cdot (\gamma s \vec{v})\\  &=\gamma \frac{\partial s}{\partial t}
  + s \frac{\partial \gamma}{\partial t} + s (\nabla \gamma \cdot
  \vec{v}) + \gamma (\nabla s \cdot \vec{v}) + \gamma s (\nabla \cdot
  \vec{v}).
\label{eq:fourdiv}
  \end{split}
 \end{equation}

The term $\gamma \frac{\partial s}{\partial t}$ reflects the local
time evolution of entropy density, while $s \frac{\partial
  \gamma}{\partial t}$ captures changes in the Lorentz factor over
time due to the motion of the system or its subsystems. In an
accelerating frame, $\gamma$ varies, and this term accounts for the
effect of velocity changes on entropy evolution. In the rest frame of
the system, where $\gamma=1$, the $s \frac{\partial \gamma}{\partial
  t}$ term does not contribute to entropy evolution. However, the
contributions from subsystems could still be relevant in presence of
their motion and these must be summed across the system. 

The spatial component, $\nabla \cdot (\gamma s \vec{v})$, represents
the spatial entropy flux arising from entropy density gradients and
velocity fields, capturing the ``spread'' of entropy throughout the
system. In non-equilibrium, this flux reflects entropy moving from
regions of higher to lower density. Analyzing its terms individually,
$s (\nabla \gamma \cdot \vec{v})$ describes changes in entropy flux
due to spatial variations in $\gamma$, $\gamma (\nabla s \cdot
\vec{v})$ relates to entropy gradients, and $\gamma s (\nabla \cdot
\vec{v})$ accounts for entropy contributions from expansions or
compressions within the system. In an accelerating frame, the $\gamma
s (\nabla \cdot \vec{v})$ term represents the divergence of the
relative velocity between the system and the observer, causing a
decrease in the entropy flux.

In equilibrium, the rest frame simplifies the entropy dynamics. The
terms related to internal gradients, such as $\nabla s$ and $\nabla
\gamma$, along with the time derivative of $\gamma$, effectively
vanish. In the thermal equilibrium, the absence of spatial variations
in entropy density or temperature reduces the four-divergence of
four-entropy to a stable expression, meaning entropy flux remains
constant over time. This aligns with the second law of thermodynamics,
as entropy neither increases nor decreases. However, if we observe the
system from a different frame moving with velocity $\vec{v}$, the
spatial flux of entropy depends on the apparent velocity gradients and
the spatial gradients of the Lorentz factor. Additionally, the time
derivative of the Lorentz factor, $\frac{\partial \gamma}{\partial
  t}=\gamma^3 \frac{\vec{v} \cdot \vec{a}}{c^2}$, also contributes to
the four-divergence of the four-entropy in the accelerating frame.\\

\noindent{\bf The Lorentz invariance of $\partial_{\mu} S^{\mu}$ and
  its non-negativity across frames}\\
In the rest frame of the system, the entropy 4-vector is defined as
$S^{\mu}=(s c,0,0,0)$, assuming that the entropy density $s$ is
uniform in space and time and there is no spatial entropy flow within
the system. The norm of the four-entropy $S^{\mu}S_{\mu}=s^2 c^2$ is a
Lorentz invariant quantity. Consequently, the four-divergence of
four-entropy, $\partial_{\mu} S^{\mu}$, is zero in the rest frame of
the system. The Lorentz invariance of $\partial_{\mu} S^{\mu}$ ensures
that it remains zero in all frames, including accelerating ones. Now
consider a frame moving with uniform acceleration $a$ relative to the
system. According to \autoref{eq:fourdiv}, the four-divergence of
four-entropy in this accelerated frame will be given by
\begin{equation}
\begin{split}
  \partial_{\mu} S^{\mu}=\frac{\partial (s \gamma)}{\partial t} +
  \nabla \cdot (\gamma s \vec{v})=s \frac{\partial \gamma}{\partial t}
  + s (\nabla \gamma \cdot \vec{v}) + \gamma s (\nabla \cdot \vec{v}).
\label{eq:accl}
\end{split}
\end{equation}
In an accelerating frame, entropy flux arises from the gradient of the
Lorentz factor, $\nabla \gamma$, and the divergence of relative
velocity, $\nabla \cdot \vec{v}$ . However, these contributions are
precisely balanced by the time derivative of the Lorentz factor,
$\frac{\partial \gamma}{\partial t}$, resulting in a net sum of
zero. This analysis also extends to systems with non-uniform entropy
density and internal velocity flows, where combined time and spatial
derivatives remain non-negative, consistent with the causal structure
of spacetime in special relativity. Relativistic causality restricts
the entropy flow direction in such a way that entropy cannot decrease
in the rest frame of the system or globally. This causality constraint
inherently connects $\partial_{\mu} S^{\mu}$ to a net non-negative
trend, affirming entropy's growth or constancy over space and time in
closed systems.

The Lorentz invariance of $\partial_{\mu} S^{\mu}$ guarantees that if
the generalized second law holds in the rest frame, it must also hold
in all inertial and accelerating frames. In a moving frame, relative
velocities may change the observed entropy flux, but the invariant
nature of the four-divergence ensures that net entropy cannot
decrease. In accelerating frames, additional terms linked to apparent
forces modify local entropy flux. But, the causal structure,
constrained by light cones, keeps $\partial_{\mu} S^{\mu}$
non-negative across all frames, thereby upholding the second law of
thermodynamics.

Thus, the generalized second law ensures that in any frame, the
combined contributions of time and spatial components to entropy yield
an overall non-negative divergence, signifying increasing disorder and
thermodynamic irreversibility.\\

\begin{figure}[htbp!]
\centering
\includegraphics[width = 0.5\textwidth]{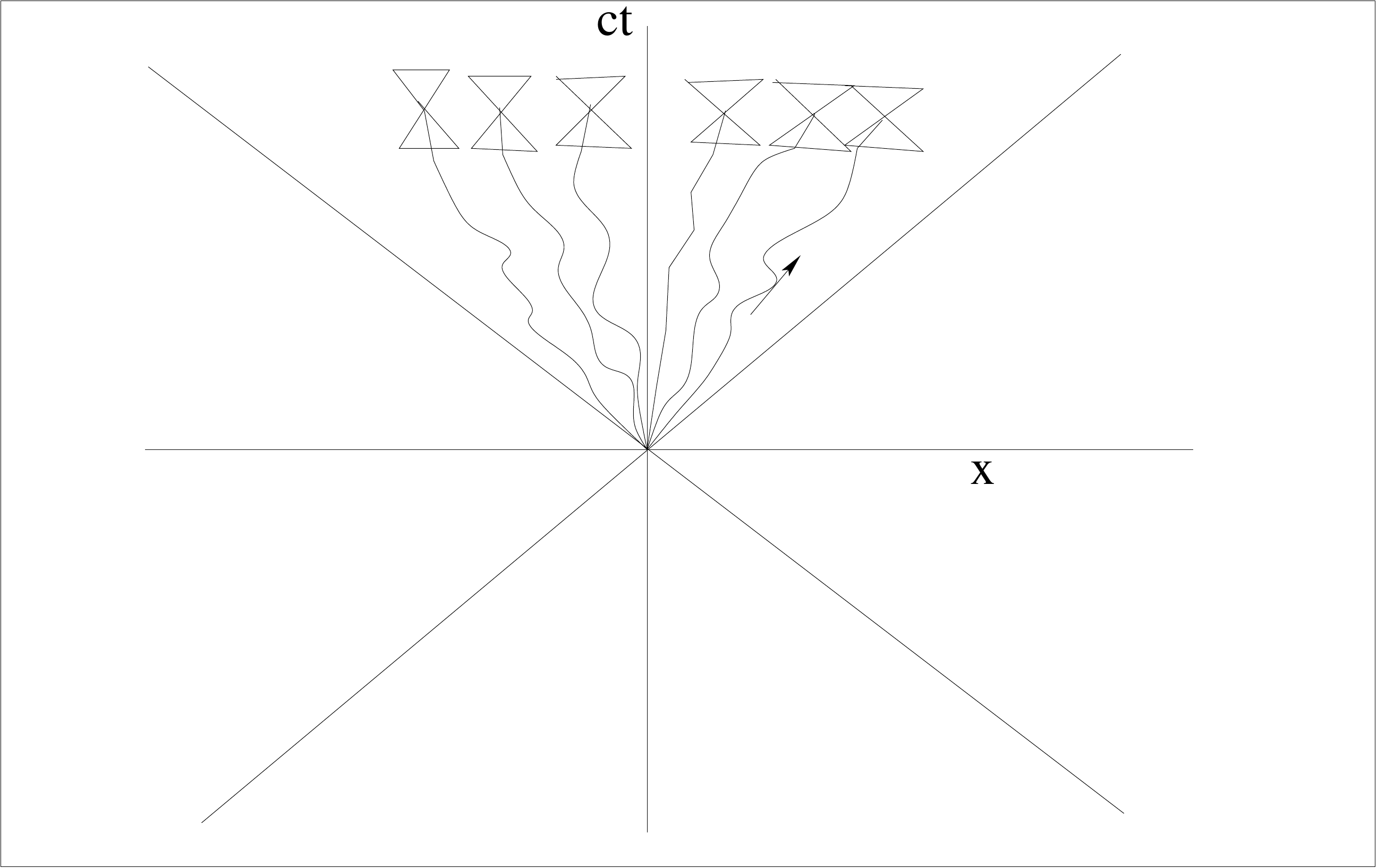}
\caption{This figure illustrates the worldlines of six perfume
  molecules dispersing through a room. The spatial spread corresponds
  to the widening of future light cones of the molecules on the
  spacetime diagram, enabling the system to access a greater number of
  microstates within causal limits. The temporal evolution reduces the
  overlap between the future light cones of the molecules, leading to
  an increase in accessible microstates. The four-entropy current of
  the system guarantees that this evolution remains irreversible. For
  clarity, only one spatial dimension and time are represented in this
  spacetime diagram.}
 \label{fig:fig1}
\end{figure}

\begin{figure}[htbp!]
\centering
\includegraphics[width = 0.5\textwidth]{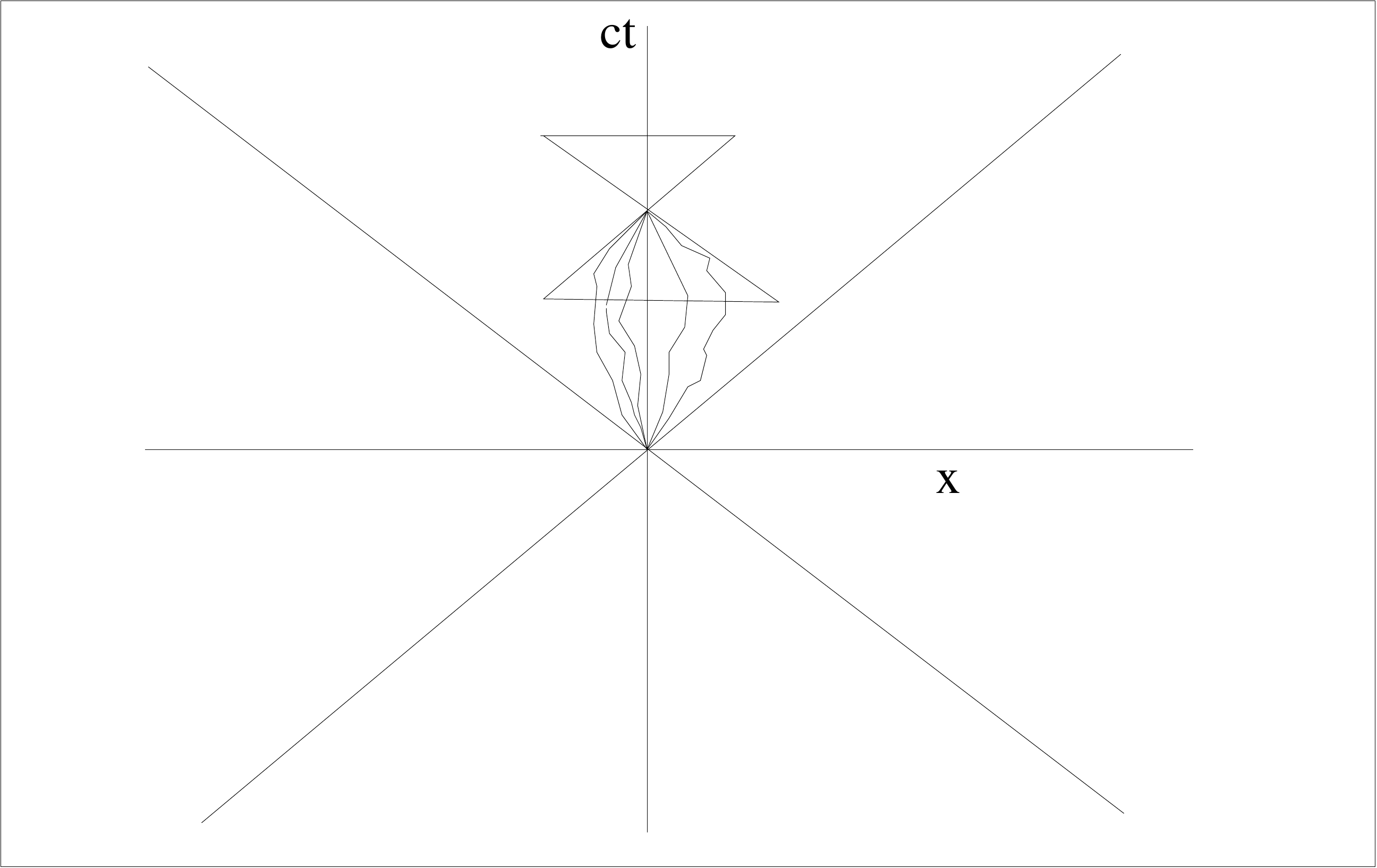}
\caption{This figure illustrates why the molecules cannot ``unspread''
  back to a more ordered initial state without violating causality. At
  any moment, the coordinates of each molecule in spacetime is
  determined solely by events within its past light cone. The
  molecules progressively spread into regions permitted by these past
  events, reducing the overlap between their past light cones.. At $c
  t>0$, their spacetime coordinates are determined by different sets
  of events in their individual past light cones. Moreover, since the
  molecules can only affect events within their future light cones,
  they can not spontaneously return to a more ordered initial
  configuration.}
 \label{fig:fig2}
\end{figure}

\noindent{\bf Causal structure of spacetime and entropy increase - An
  intuitive example}\\
In special relativity, spacetime structure is defined by the light
cone, which establishes causal relationships between events. Each
particle or system has a light cone that determines the regions of
spacetime it can influence (the future light cone) and those that can
influence it (the past light cone). These light cones ensure that
effects follow causes, preserving the order of events and preventing
backward causality. This causal structure shapes not only particle
interactions but also the flow and distribution of entropy in
spacetime.

We present an intuitive example to illustrate how the causal structure
of spacetime enforces the second law of thermodynamics. Imagine
perfume molecules spreading through a room. Initially, the molecules
are confined near the nozzle, with similar positions and limited
diversity in velocity and direction. However, as they start moving and
interacting, their trajectories diverge due to tiny velocity
differences and collisions. Each molecule's movement is constrained by
its light cone, indicating it can only influence events within its own
future light cone. The molecules begin with nearly identical past
light cones, but as they spread, their forward light cones define
causally bound regions they can reach. Over time, this accessible
spatial region expands, increasing the number of possible microstates
and disorder (\autoref{fig:fig1}).

The causal structure ensures that once dispersed, the molecules cannot
spontaneously gather back at the nozzle. Each molecule's position and
movement are tied to its previous trajectory, constrained by its light
cone, preventing a return to the initial state
(\autoref{fig:fig2}). They continuously move into regions allowed by
their past positions, unable to reverse to a more ordered
configuration due to the irreversibility embedded in the causal
structure. This constraint drives molecules toward larger volumes with
more configurations, increasing entropy. Thus, the evolution of the
system towards equilibrium aligns naturally with entropy increase, as
the accessible spacetime regions fill in with time, reinforcing the
second law.

Entropy increases or remains constant over time, defining the arrow of
time. This irreversible progression stems from the fact that future
events are only accessible from past states if causality is
respected. As the system evolves, entropy spreads spatially which can
be linked to processes like heat transfer or particle movement. This
spatial spreading (related to entropy flux) corresponds to the
widening of the future light cone on a spacetime diagram. The
expanding future light cones associated with the molecules allow the
system to access more microstates, within the bounds of causality. The
growth of accessible microstates also reflects the temporal evolution
(related to time derivative of entropy) of the system, while the
four-entropy current ensures that this evolution is
irreversible. Thus, the second law of thermodynamics is not incidental
but deeply tied to the structure of spacetime in special
relativity.\\ \\
\noindent{\bf Implications}\\
The four-entropy $S^{\mu}$ extends the concept of scalar entropy to
relativistic contexts, ensuring that the second law holds across all
frames. Since special relativity prohibits information or causal
influence from exceeding the speed of light, irreversible processes
such as collisions and heat flow, can only move forward in
time. Consequently, entropy must either increase or remain constant,
ensuring $\partial_{\mu} S^{\mu} \geq 0$. The Lorentz invariance of
the four-divergence of four-entropy links entropy to the geometry of
spacetime, indicating that the arrow of time and the irreversible
nature of thermodynamic processes stem directly from the structure of
special relativity. Entropy production thus respects causal structure
of spacetime, reinforcing that the second law of thermodynamics aligns
with causality. This framework offers a coherent and unified
explanation of the second law, extending beyond individual frames to
account for the relativistic behaviour of entropy in both inertial and
accelerating frames.

The proposed framework offers promising insights for curved
spacetimes, such as those in general relativity, where the second law
of thermodynamics also dictates a forward arrow of time. In this
context, entropy evolves according to the covariant derivative
$\nabla_{\mu} S^{\mu} = \partial_{\mu} S^{\mu}+\Gamma_{\mu\nu}^{\mu}
S^{\nu} \geq 0$, maintaining the second law within curved
spacetime. Here, $\Gamma_{\mu\nu}^{\mu}$ are the Christofell symbols,
encoding the effects of spacetime curvature. In the presence of
gravity, the second law of thermodynamics is deeply interwoven with
the curvature of spacetime. Extending four-entropy into the realm of
general relativity could deepen our understanding of thermodynamics in
extreme conditions such as near black hole event horizons or in the
early Universe. Moreover, the expansion and acceleration of the
universe could be tied to the second law, ensuring that entropy
increases on cosmological scales \citep{pandey17}. Exploring this
issue through the evolution of four-entropy offers an exciting avenue
for further study.

The second law is deeply rooted in the fundamental structure of
spacetime. The generalized second law within the four-vector framework
offers a fresh perspective for examining thermodynamic principles
across a broad spectrum of phenomena.


\begin{thebibliography}{44}%
	\makeatletter
	\providecommand \@ifxundefined [1]{%
		\@ifx{#1\undefined}
	}%
	\providecommand \@ifnum [1]{%
		\ifnum #1\expandafter \@firstoftwo
		\else \expandafter \@secondoftwo
		\fi
	}%
	\providecommand \@ifx [1]{%
		\ifx #1\expandafter \@firstoftwo
		\else \expandafter \@secondoftwo
		\fi
	}%
	\providecommand \natexlab [1]{#1}%
	\providecommand \enquote  [1]{``#1''}%
	\providecommand \bibnamefont  [1]{#1}%
	\providecommand \bibfnamefont [1]{#1}%
	\providecommand \citenamefont [1]{#1}%
	\providecommand \href@noop [0]{\@secondoftwo}%
	\providecommand \href [0]{\begingroup \@sanitize@url \@href}%
	\providecommand \@href[1]{\@@startlink{#1}\@@href}%
	\providecommand \@@href[1]{\endgroup#1\@@endlink}%
	\providecommand \@sanitize@url [0]{\catcode `\\12\catcode `\$12\catcode
		`\&12\catcode `\#12\catcode `\^12\catcode `\_12\catcode `\%12\relax}%
	\providecommand \@@startlink[1]{}%
	\providecommand \@@endlink[0]{}%
	\providecommand \url  [0]{\begingroup\@sanitize@url \@url }%
	\providecommand \@url [1]{\endgroup\@href {#1}{\urlprefix }}%
	\providecommand \urlprefix  [0]{URL }%
	\providecommand \Eprint [0]{\href }%
	\providecommand \doibase [0]{http://dx.doi.org/}%
	\providecommand \selectlanguage [0]{\@gobble}%
	\providecommand \bibinfo  [0]{\@secondoftwo}%
	\providecommand \bibfield  [0]{\@secondoftwo}%
	\providecommand \translation [1]{[#1]}%
	\providecommand \BibitemOpen [0]{}%
	\providecommand \bibitemStop [0]{}%
	\providecommand \bibitemNoStop [0]{.\EOS\space}%
	\providecommand \EOS [0]{\spacefactor3000\relax}%
	\providecommand \BibitemShut  [1]{\csname bibitem#1\endcsname}%
	\let\auto@bib@innerbib\@empty

        
	\bibitem [{\citenamefont {Einstein}(1907)}]{einstein07}%
	\BibitemOpen
	\bibfield  {author} {\bibinfo {author} {\bibfnamefont {Albert}\ \bibnamefont
			{Einstein}},\ }\bibfield  {title} {\enquote {\bibinfo {title} {{Über das Relativitätsprinzip und die aus demselben gezogenen Folgerungen}},}\ }\href {\doibase} {\bibfield
		{journal} {\bibinfo  {journal} {Jahrbuch der Radioaktivität und Elektronik}\ }\textbf {\bibinfo {volume} {4}},\
		\bibinfo {pages} {411} (\bibinfo {year} {1907})}\BibitemShut {NoStop}%
      \bibitem [{\citenamefont {Planck}(1908)}]{planck08}%
        \BibitemOpen
	\bibfield  {author} {\bibinfo {author} {\bibfnamefont {Max}\ \bibnamefont
			{Planck}},\ }\bibfield  {title} {\enquote {\bibinfo {title} {{Zur Dynamik bewegter Systeme}},}\ }\href {\doibase 10.1002/andp.19083310602} {\bibfield
		{journal} {\bibinfo  {journal} {Annalen Der Physik. (Leipzig)}\ }\textbf {\bibinfo {volume} {26}},\
		\bibinfo {pages} {1} (\bibinfo {year} {1908})}\BibitemShut {NoStop}%
      \bibitem [{\citenamefont {Ott}(1963)}]{ott63}%
        \BibitemOpen
	\bibfield  {author} {\bibinfo {author} {\bibfnamefont {Heinrich}\ \bibnamefont
			{Ott}},\ }\bibfield  {title} {\enquote {\bibinfo {title} {{Lorentz-Transformation der Wa\"rme und der Temperatur}},}\ }\href {\doibase 10.1007/BF01375397} {\bibfield
		{journal} {\bibinfo  {journal} {Zeitschrift fu\"r Physik}\ }\textbf {\bibinfo {volume} {175}},\
		\bibinfo {pages} {70} (\bibinfo {year} {1963})}\BibitemShut {NoStop}%
      \bibitem [{\citenamefont {Arzeliès}(1965)}]{arzelies65}%
        \BibitemOpen
	\bibfield  {author} {\bibinfo {author} {\bibfnamefont {Henri}\ \bibnamefont
			{Arzeliès}},\ }\bibfield  {title} {\enquote {\bibinfo {title} {{Transformation relativiste de la température et de quelques autres grandeurs thermodynamiques}},}\ }\href {\doibase 10.1007/BF02739342} {\bibfield	{journal} {\bibinfo  {journal} {Il Nuovo Cimento}\ }\textbf {\bibinfo {volume} {35}},\
		\bibinfo {pages} {792} (\bibinfo {year} {1965})}\BibitemShut {NoStop}%
      \bibitem [{\citenamefont {Landsberg}(1967)}]{landsberg67}%
        \BibitemOpen
	\bibfield  {author} {\bibinfo {author} {\bibfnamefont {Peter Theodore}\ \bibnamefont
			{Landsberg}},\ }\bibfield  {title} {\enquote {\bibinfo {title} {{Does a Moving Body Appear Cool?}},}\ }\href {\doibase 10.1038/214903a0} {\bibfield
		{journal} {\bibinfo  {journal} {Nature}\ }\textbf {\bibinfo {volume} {214}},\
		\bibinfo {pages} {903} (\bibinfo {year} {1967})}\BibitemShut {NoStop}%
      \bibitem [{\citenamefont {Landsberg}\ and\ \citenamefont {Johns}(1967)}]{landsbergJohns67}%
        \BibitemOpen
	\bibfield  {author} {\bibinfo {author} {\bibfnamefont {P.T.}\ \bibnamefont
			{Landsberg}} \ and\ \bibinfo {author} {\bibfnamefont {K.A.}\ \bibnamefont {Johns}}},\bibfield  {title} {\enquote {\bibinfo {title} {{A Relativistic Generalization of Thermodynamics}},}\ }\href {\doibase 10.1007/BF02710651} {\bibfield
		{journal} {\bibinfo  {journal} {Il Nuovo Cimento B}\ }\textbf {\bibinfo {volume} {52}},\
		\bibinfo {pages} {28} (\bibinfo {year} {1967})}\BibitemShut {NoStop}%
 	\bibitem [{\citenamefont {Sackur}(1912)}]{sackur12}%
	\BibitemOpen
	\bibfield  {author} {\bibinfo {author} {\bibfnamefont {Otto}\ \bibnamefont
			{Sackur}},\ }\bibfield  {title} {\enquote {\bibinfo {title} {{Die Bedeutung des elementaren Wirkungsquantums f\"ur die Gastheorie und die Berechnung der chemischen Konstanten}},}\ }\href {\doibase} {\bibfield {journal} {\bibinfo  {journal} {Festschrift W. Nernst zu seinem 25j\"ahrigen Doktorjubil\"aum}\ }\textbf {\bibinfo {volume} {}},\ \bibinfo {pages} {405} (\bibinfo {year} {1912})}\BibitemShut {NoStop}%
      \bibitem [{\citenamefont {Tetrode}(1912)}]{tetrode12}%
        \BibitemOpen
	\bibfield  {author} {\bibinfo {author} {\bibfnamefont {Hugo}\ \bibnamefont
			{Tetrode}},\ }\bibfield  {title} {\enquote {\bibinfo {title} {{Die chemische Konstante der Gase und das elementare Wirkungsquantum}},}\ }\href {\doibase} {\bibfield
		{journal} {\bibinfo  {journal} {Annalen Der Physik}\ }\textbf {\bibinfo {volume} {38}},\
		\bibinfo {pages} {434} (\bibinfo {year} {1912})}\BibitemShut {NoStop}%
      \bibitem [{\citenamefont {Taff}(1968)}]{taff68}%
        \BibitemOpen
	\bibfield  {author} {\bibinfo {author} {\bibfnamefont {L. G.}\ \bibnamefont
			{Taff}},\ }\bibfield  {title} {\enquote {\bibinfo {title} {{A relativistic transformation for temperature}},}\ }\href {\doibase 10.1016/0375-9601(68)90074-1} {\bibfield
		{journal} {\bibinfo  {journal} {Physics Letters A}\ }\textbf {\bibinfo {volume} {27}},\
		\bibinfo {pages} {605} (\bibinfo {year} {1968})}\BibitemShut {NoStop}%
        \bibitem [{\citenamefont {Dantzig}(1939)}]{dantzig39}%
        \BibitemOpen
	\bibfield  {author} {\bibinfo {author} {\bibfnamefont {D. van}\ \bibnamefont
			{Dantzig}},\ }\bibfield  {title} {\enquote {\bibinfo {title} {{On the phenomenological thermodynamics of moving matter}},}\ }\href {\doibase 10.1016/S0031-8914(39)90072-8} {\bibfield
		{journal} {\bibinfo  {journal} {Physica}\ }\textbf {\bibinfo {volume} {6}},\
		\bibinfo {pages} {673} (\bibinfo {year} {1939})}\BibitemShut {NoStop}%
      \bibitem [{\citenamefont {Kampen}(1968)}]{vankampen68}%
        \BibitemOpen
	\bibfield  {author} {\bibinfo {author} {\bibfnamefont {N. G. van}\ \bibnamefont
			{Kampen}},\ }\bibfield  {title} {\enquote {\bibinfo {title} {{Relativistic Thermodynamics of Moving Systems}},}\ }\href {\doibase 10.1103/PhysRev.173.295} {\bibfield
		{journal} {\bibinfo  {journal} {Phys. Rev.}\ }\textbf {\bibinfo {volume} {173}},\
		\bibinfo {pages} {295} (\bibinfo {year} {1968})}\BibitemShut {NoStop}%
      \bibitem [{\citenamefont {Israel}(1981)}]{israel81}%
        \BibitemOpen
	\bibfield  {author} {\bibinfo {author} {\bibfnamefont {W.}\ \bibnamefont
			{Israel}},\ }\bibfield  {title} {\enquote {\bibinfo {title} {{Thermodynamics of relativistic systems}},}\ }\href {\doibase 10.1016/0378-4371(81)90220-X} {\bibfield
		{journal} {\bibinfo  {journal} {Physica A}\ }\textbf {\bibinfo {volume} {106}},\
		\bibinfo {pages} {204} (\bibinfo {year} {1981})}\BibitemShut {NoStop}%
         \bibitem [{\citenamefont {Nakamura}(2006)}]{nakamura06}%
        \BibitemOpen
	\bibfield  {author} {\bibinfo {author} {\bibfnamefont {Tadas K.}\ \bibnamefont
			{Nakamura}},\ }\bibfield  {title} {\enquote {\bibinfo {title} {{Covariant thermodynamics of an object with finite volume}},}\ }\href {\doibase 10.1016/j.physleta.2005.11.070} {\bibfield
		{journal} {\bibinfo  {journal} {Physics Letters A}\ }\textbf {\bibinfo {volume} {352}},\
		\bibinfo {pages} {175} (\bibinfo {year} {2006})}\BibitemShut {NoStop}%
        \bibitem [{\citenamefont {Wu}(2009)}]{wu09}%
        \BibitemOpen
	\bibfield  {author} {\bibinfo {author} {\bibfnamefont {Jhong Chao}\ \bibnamefont
			{Wu}},\ }\bibfield  {title} {\enquote {\bibinfo {title} {{Inverse-temperature 4-vector in special relativity}},}\ }\href {\doibase 10.1209/0295-5075/88/20005} {\bibfield
		{journal} {\bibinfo  {journal} {Europhysics Letters}\ }\textbf {\bibinfo {volume} {88}},\
		\bibinfo {pages} {20005} (\bibinfo {year} {2009})}\BibitemShut {NoStop}%
         \bibitem [{\citenamefont {Davies}(1975)}]{davies75}%
        \BibitemOpen
	\bibfield  {author} {\bibinfo {author} {\bibfnamefont {P. C. W.}\ \bibnamefont
			{Davies}},\ }\bibfield  {title} {\enquote {\bibinfo {title} {{Scalar production in Schwarzschild and Rindler metrics}},}\ }\href {\doibase 10.1088/0305-4470/8/4/022} {\bibfield
		{journal} {\bibinfo  {journal} {Journal of Physics A: Mathematical and General}\ }\textbf {\bibinfo {volume} {8}},\
		\bibinfo {pages} {609} (\bibinfo {year} {1975})}\BibitemShut {NoStop}%
         \bibitem [{\citenamefont {Unruh}(1976)}]{unruh76}%
        \BibitemOpen
	\bibfield  {author} {\bibinfo {author} {\bibfnamefont {W.G.}\ \bibnamefont
			{Unruh}},\ }\bibfield  {title} {\enquote {\bibinfo {title} {{Notes on black-hole evaporation}},}\ }\href {\doibase 10.1103/PhysRevD.14.870} {\bibfield
		{journal} {\bibinfo  {journal} {Phys. Rev. D}\ }\textbf {\bibinfo {volume} {14}},\
		\bibinfo {pages} {870} (\bibinfo {year} {1976})}\BibitemShut {NoStop}%
      \bibitem [{\citenamefont {Pandey}(2024)}]{pandey17}%
        \BibitemOpen
	\bibfield  {author} {\bibinfo {author} {\bibfnamefont {Biswajit}\ \bibnamefont
	    {Pandey}},\ }\bibfield  {title} {\enquote {\bibinfo {title} {{Does information entropy play a role in the expansion and acceleration of the Universe?}},}\ }\href {\doibase
          10.1093/mnrasl/slx109} {\bibfield
		{journal} {\bibinfo  {journal} {MNRAS Letters}\ }\textbf {\bibinfo {volume} {471}},\
		\bibinfo {pages} {L77} (\bibinfo {year} {2017})}\BibitemShut {NoStop}%

\end{thebibliography}
\end{document}